\documentclass[conference]{IEEEtran}

\usepackage{tabularx}
\usepackage{amsthm}
\usepackage{epstopdf}
\usepackage{cite}
\usepackage{bbm}
\usepackage{amsmath,amssymb,amsfonts}
\usepackage{algorithm} 
\usepackage{algpseudocode} 
\usepackage{graphicx}
\usepackage{textcomp}
\usepackage{todonotes}
\usepackage{xcolor}
\usepackage[utf8]{inputenc} 
\usepackage[T1]{fontenc}    
\usepackage{amsfonts}       
\usepackage{nicefrac}       
\usepackage{microtype}      
\usepackage{wrapfig}
\usepackage{color}
\usepackage{float}
\usepackage{listings}
\usepackage{amsmath}
\usepackage{graphicx}
\usepackage{makecell}
\usepackage{tabularx}
\usepackage{caption}%
\usepackage{nopageno}
\usepackage{subcaption}
\usepackage{amsmath}
\usepackage{amssymb}
\makeatletter
\def\ps@headings{%
\def\@oddhead{\mbox{}\scriptsize\rightmark \hfil \thepage}%
\def\@evenhead{\scriptsize\thepage \hfil \leftmark\mbox{}}%
\def\@oddfoot{}%
\def\@evenfoot{}}
\makeatother
\usepackage{caption}
\usepackage[letterpaper, left=0.641in, right=0.641in, bottom=1.095in, top=0.73in]{geometry}
\newcolumntype{A}{>{\centering\arraybackslash}p{0.3\textwidth}}
\newcolumntype{B}{>{\centering\arraybackslash}p{0.17\textwidth}}
\usepackage[compact]{titlesec}
    \titlespacing{\section}{0pt}{0.3ex}{0.3ex}
    \titlespacing{\subsection}{0pt}{0.2ex}{0.2ex}
    \titlespacing{\subsubsection}{0pt}{0.2ex}{0.2ex}

\usepackage{framed}
\usepackage{enumitem}
\everymath{\small}
\everydisplay{\small}

\begin{document}

\title{
\textsc{SliceWise}: Traffic-Aware Multi-Agent RL with Temporal Encoding for Joint Slice Admission and Resource Orchestration in O-RAN \vspace{-0.1in}
}

\author{Xingqi Wu$^{\dagger}$, Yuhui Wang$^{\dagger}$, Junaid Farooq$^{\dagger}$, Hakim Ghazzai$^{*}$, and Gianluca Setti$^{*}$\\
$^\dagger$Department of Electrical and Computer Engineering, University of Michigan-Dearborn, MI USA,\\
$^*$King Abdullah University of Science \& Technology (KAUST), Thuwal, Makkah, Saudi Arabia.\\
Emails: \{xingqiwu, ywangdq, mjfarooq\}@umich.edu, \{hakim.ghazzai, gianluca.setti\}@kaust.edu.sa.\vspace{-0.2in}
}

\maketitle

\begin{abstract}

Ultra-reliable low-latency communication (URLLC) services in mission-critical applications such as industrial automation, remote healthcare, and autonomous systems demand stringent quality-of-service (QoS) guarantees under dynamic, resource-constrained wireless conditions. Open radio access networks (O-RAN) offer a software-defined and programmable foundation for embedding AI-driven control. This letter proposes \textsc{SliceWise}, a traffic-aware multi-agent deep reinforcement learning (MARL) framework for joint admission control, user-slice association, and resource orchestration in O-RAN. The control problem is modeled as a cooperative multi-agent Markov decision process, with long short-term memory (LSTM) networks capturing temporal traffic dynamics and dueling double deep Q-networks (D3QN) ensuring stable learning. Simulation results show that \textsc{SliceWise} substantially reduces delay violations and packet drops while improving slice-level reliability and resource efficiency, offering a scalable orchestration solution for AI-native O-RAN deployments.
\end{abstract}


\IEEEpeerreviewmaketitle
\vspace{-0.0in}
\section{Introduction}
The proliferation of latency-sensitive and mission-critical applications in 5G and beyond, including industrial automation, remote healthcare, and connected autonomous systems, imposes stringent quality-of-service (QoS) requirements on radio access networks (RAN)~\cite{ns-magazine}. Ultra-reliable low-latency communication (URLLC) demands millisecond-level end-to-end latency and extremely high reliability under dynamic and bursty traffic conditions~\cite{bursty}. Ensuring these guarantees in the presence of limited spectrum resources, heterogeneous service demands, and fluctuating network loads is challenging, particularly in network slicing environments where maintaining service-level agreements (SLAs) across slices during congestion is complex.  

Open RAN (O-RAN) provides a disaggregated, software-defined architecture with programmable interfaces that enable fine-grained control of heterogeneous services~\cite{oran-arch}. The near-real-time RAN Intelligent Controller (near-RT RIC) facilitates data-driven decision-making for admission control, user-slice association, and resource orchestration~\cite{NT-RIC}. While these capabilities position O-RAN as an ideal platform for embedding AI-driven control, existing resource and admission control mechanisms remain inadequate for URLLC. Static bandwidth partitioning, heuristic scheduling, and centralized optimization~\cite{convex-slicing} often lack adaptability to bursty traffic, and recent machine learning (ML)~\cite{ATHENA} and reinforcement learning (RL)~\cite{SLA_RL, flexible-ran-slicing} approaches typically treat admission control, slice association, and resource orchestration in isolation, neglecting their interdependence and the distributed nature of O-RAN.  
\begin{figure}[t!]
    \centering
    \vspace{-0.in}
    \includegraphics[width = 3.5in]{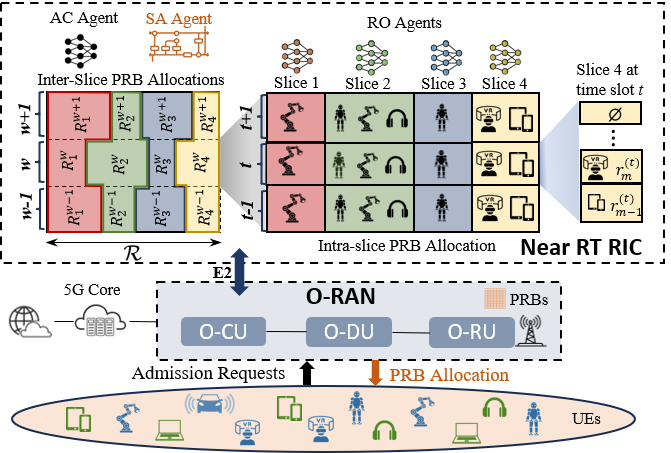}
    \caption{\small Integrated admission control and resource orchestration framework in O-RAN architecture. \vspace{-0in}}
    \label{fig:workflows}
\end{figure}

This letter proposes \textsc{SliceWise}, a traffic-aware multi-agent deep reinforcement learning (MARL) framework for joint admission control, slice association, and resource orchestration in O-RAN. The problem is modeled as a cooperative multi-agent Markov decision process (MMDP), with agents specialized for each control function. Long short-term memory (LSTM) networks capture temporal traffic dynamics, and dueling double deep Q-networks (D3QN) provide stable and efficient policy learning in high-dimensional, non-stationary environments.  
The main contributions are as follows:  
\begin{enumerate}
    \item We propose a unified MARL framework that jointly optimizes admission control, slice association, and resource allocation for O-RAN;
    \item We develop LSTM-based agent designs that exploit temporal features and balance URLLC reliability with resource efficiency;
    \item Extensive simulations show that \textsc{SliceWise} improves delay, reliability, and efficiency compared to random, empirical, and partially coordinated RL baselines.
\end{enumerate}

\vspace{-0.0in}
\section{System Model}\label{section2}

We consider an O-RAN architecture supporting network slicing for differentiated service provisioning as illustrated in Fig.~\ref{fig:workflows}. 
Let $\mathcal{M}=\{1,2,\dots, M\}$ denote the set of user equipments (UEs) served by an O-RAN radio unit (O-RU) under a centralized unit (O-CU), which coordinates with the distributed unit (O-DU) to allocate $\mathcal{R} \in \mathbb{Z}^+$ physical resource blocks (PRBs). 

\subsection{Inter-Slice Resource Partitioning and Admission Control}

The system operates over a finite time horizon $\mathcal{T} = \{0, 1, \dots, T-1\}$, with each time slot $t \in \mathcal{T}$ of fixed duration $T_s$. To support dynamic and SLA-driven traffic management, the RAN is logically partitioned into a set of active slices $\mathcal{S} = \{1, 2, \dots, S\}$, each designed to serve a specific class of traffic. Resource allocation decisions are made at discrete reconfiguration windows, indexed by $w \in \mathcal{W} = \{0, 1, \dots, \left\lfloor \frac{T}{T_w} \right\rfloor - 1 \}$, where $T_w$ denotes the duration (in slots) of each window. The $w^{\text{th}}$ reconfiguration window starts at time $t_w$ and spans the time slots $\mathcal{T}_w = \{t_w, t_w + 1, \dots, t_w + T_w - 1\}$, with $t_{w+1} = t_w + T_w$ to ensure non-overlapping intervals.
At the beginning of each window, the near-RT RIC performs user admission control and inter-slice PRB partitioning. Each slice $s \in \mathcal{S}$ is assigned a resource budget $R_s^w \in \mathbb{Z}^+$, constrained by the total PRBs, i.e., $\sum_{s \in \mathcal{S}} R_s^w \leq \mathcal{R}, \forall w \in \mathcal{W}$.
Let $\mathcal{M}_a^{t_w}$ denote the set of UEs requesting admission at the start of window $w$, and $\mathcal{M}_b^{t_w}$ denote the currently admitted users, with $\mathcal{M}_a^{t_w} \cap \mathcal{M}_b^{t_w} = \emptyset$ and $\mathcal{M}_a^{t_w} \cup \mathcal{M}_b^{t_w} = \mathcal{M}$. The binary admission decision for UE $m \in \mathcal{M}_a^{t_w}$ is represented as $a_m^w \in \{0,1\}$, where 1 denotes admission. The complete admission vector is defined as
$
\mathbf{a}_w = \left[a_1^w, a_2^w, \dots, a_{\lvert \mathcal{M}_a^{t_w}\rvert}^w\right].
$  
The set of newly admitted UEs is denoted as $\mathcal{M}_k^{t_w} = \{ m \in \mathcal{M}_a^{w} \mid a_m^w = 1 \}$, and its cardinality is defined as \(M_k^{t_w} = |\mathcal{M}^{t_w}_k|\).
The set of active UEs is then updated as $\mathcal{M}_b^{t_w} = \mathcal{M}_b^{t_{w-1}} \cup \mathcal{M}_k^{t_w}$, and remains unchanged throughout the set of time slots $t \in \mathcal{T}_w$, i.e., $\mathcal{M}_b^t = \mathcal{M}_b^{t_w}$.
User-slice association is determined by a binary matrix
$
\mathbf{Z}_w = \left[\mathbf{z}_{1}^w, \mathbf{z}_{2}^w, \dots, \mathbf{z}_{M_b^{t_w}}^w\right],
$
where $\mathbf{z}_m^w = [z_{m,1}^w, z_{m,2}^w, \dots, z_{m,S}^w]^{\text{T}}$ assigns UE $m$ to exactly one slice, i.e., $\sum_{s \in \mathcal{S}} z_{m,s}^w = 1, \forall w$. Accordingly, the UEs associated with slice $s$ at window $w$ are defined as $\mathcal{M}_s^{t_w} = \{ m \in \mathcal{M}_b^{t_w} \mid z_{m,s}^w = 1 \}$ with $\bigcup_{s\in \mathcal{S}} \mathcal{M}^{t_w}_s = \mathcal{M}^{t_w}_b$.

\subsection{Intra-Slice Resource Allocation and Service Reliability}

Each UE \( m \in \mathcal{M} \) is located at a time-varying distance \( d_m(t) \) from the O-RU at time slot \( t \in \mathcal{T}\). The average path loss $\beta_m(t)$ for UE \( m \) at time \( t \) is modeled as a weighted combination of line-of-sight (LoS) and non-LoS components~\cite{3gpp-path} as    $\beta_m(t) = p \cdot \operatorname{FSPL}_{m}(t) + (1 - p) \cdot \operatorname{PL}_{m}(t)$, 
where \( p \in [0,1] \) is the LoS probability. The free-space path loss (FSPL) and non-LoS path loss (PL) components are computed as follows:  
\begin{align}  
\small
    &\operatorname{FSPL}_{m}(t) = 20 \log_{10} \left(d_{m}(t)\right) + 20 \log_{10} \left(f_c\right) - 27.55 + \eta, \\  
    &\operatorname{PL}_{m}(t) = 20 \log_{10} \left(d_{m}(t)\right) + 20 \log_{10} \left(f_c\right) - 27.55 + \psi_\xi,
\end{align}  
\normalsize
where \( f_c \) is the carrier frequency, and \( \eta \), \( \psi_\xi \) represent the excess attenuation for LoS and non-LoS links, respectively.  
Let \( \tau^\text{t}_m(t) \) denote the end-to-end transmission delay for UE \( m \) at time $t$, which is computed as follows:  
\begin{align}  
\small
    \tau^\text{t}_m(t) = \frac{\delta_m(t) N_m(t) + \left(\tau^\text{t}_m(t-1) - T_s\right)^{+} b_m(t-1)}{b_m(t)},
\end{align}  
\normalsize
where $\delta_m(t)$ is the packet size, $N_m(t)$ is the number of packets, and $b_m(t)$ is the achievable data rate given by:
\begin{align}  
\small
    b_m(t) = r_m(t)B \log_2 \left(1 + \operatorname{SINR}_m(t)\right),
\end{align}  
\normalsize
with signal-to-interference-plus-noise ratio (SINR) for UE \( m \) at time \( t \) defined as
$\operatorname{SINR}_m(t) = \frac{P 10^{-\beta_m(t)/10} |h_m(t)|^2}{r_m(t) B N_0 + I_m(t)}$, where \( P \) denotes the transmit power of the O-RU, \( r_m(t) \) is the number of PRBs allocated to UE $m$, $B$ is the bandwidth per PRB, \( |h_m(t)|^2 \) is the fading channel gain, \( N_0 \) is the noise power spectral density, and \( I_m(t) \) is the interference power. 
The intra-slice resource constraint ensures that $\sum_{m\in \mathcal{M}_s^{t}}r_m(t)\leq R_s^w$ for each slice $s$ and time $t \in \mathcal{T}_w$.
To meet SLA requirements, each UE must satisfy the reliability condition defined as 
$
\mathbb{P}\left(\tau^\text{t}_m(t) \leq \tau_m^{\max}\right) \geq \lambda,  \quad \forall m \in \mathcal{M}_b^t,
$
where \( \tau_m^{\max} \in \{\tau_{\text{e}}, \tau_{\text{u}}\} \) corresponds to latency bounds for enhanced multimedia broadband (eMBB) (\( \tau_{\text{e}} \)) and URLLC (\( \tau_{\text{u}} \)) services, and \( \lambda \in (0,1)\) is the fulfillment threshold.

\subsection{Problem Formulation}

Slice reliability measures the fraction of admitted UEs whose latency remains within SLA thresholds. It is defined at window 
$w$ for slice $s \in \mathcal{S}$ as:
\begin{align}
\small
\Omega_s\hspace{-0.03in}\left(\mathbf{a}_w, r_m(t), \mathbf{Z}_w, R_s^w\right) \hspace{-0.03in}= \hspace{-0.12in}\prod_{m \in \mathcal{M}_{s}^{t}} \hspace{-0.1in}g_s^r \left(\mathbb{P}\left(\tau_m^{\text{t}}(t) \leq \tau_m^{\max}\right) - \lambda \right), \label{eq:reliability}
\end{align}
\normalsize
where $\mathcal{M}_{s}^{t} = \mathcal{M}_{s}^{t_w}$ for all $t\in \mathcal{T}_w$ and $g_s^r(\cdot): \mathbb{R} \rightarrow [0,1]$ is a smooth, monotonically increasing function quantifying the satisfaction margin of the latency requirement defined as 
$
g_s^r(x) = \frac{1}{1 + \exp\left(-\alpha_s^r( x + \varphi)\right)},
$
where $\alpha_s^r$ controls the slope and $\varphi$ adjusts the offset to reflect sensitivity to slice-specific reliability.
To ensure temporal reliability, we define a long-term utility that combines instantaneous and discounted future performance, jointly maximizing user admission and slice reliability at time~$t$:
\begin{align}
\small
&\hspace{-0.02in}F_1\left(\mathbf{a}_w, r_m(t), \mathbf{Z}_w, R_s^w\right)=\frac{M_k^{t_w}}{M_a^{t_w}} \prod_{s\in\mathcal{S}}\Omega_s\left(\mathbf{a}_w, r_m(t), \mathbf{Z}_w, R_s^w\right)+ \notag\\
&\mathbb{E}\hspace{-0.01in}\left(\sum_{t'=1}^{\infty} \zeta^{t'}\Gamma\hspace{-0.01in}\left(\mathbf{a}_{w+\lfloor \frac{t'}{T_w}\rfloor}, r_m(t+t'), \mathbf{Z}_{w+\lfloor \frac{t'}{T_w}\rfloor}, \hspace{-0.025in}R_s^{w+\lfloor \frac{t'}{T_w}\rfloor},\mathcal{S}\hspace{-0.01in}\right)\hspace{-0.01in} \right)\hspace{-0.02in}, \label{obj-2} 
\end{align}
\normalsize
where $\zeta \in (0,1)$ is a discount factor and $\Gamma(.)$ represents the product of admission and slice reliabilities across future windows, expressed as
$\Gamma =\frac{M_k^{t_w+\lfloor \frac{t'}{T_w}\rfloor T_w}}{M_a^{t_w+\lfloor \frac{t'}{T_w}\rfloor T_w}}\underset{s\in\mathcal{S}}{\prod}\Omega_s (.)$, $t \in \mathcal{T}_w$.
A penalty term captures current and future PRB utilization across admitted users to enforce resource efficiency:
\begin{align}
\small
&\hspace{-0.15in}F_2\left(r_m(t)\right)\hspace{-0.025in}=\hspace{-0.025in}\frac{1}{\mathcal{R}}\hspace{-0.05in}\sum_{m \in \mathcal{M}_b^t} \hspace{-0.05in}r_m(t)+\mathbb{E}\left(\sum_{t'=1}^{\infty} \zeta^{t'} \frac{1}{\mathcal{R}}\hspace{-0.1in}\sum_{m \in \mathcal{M}_b^{t'}} \hspace{-0.05in}r_m(t+t')\right),\label{obj-1}
\end{align}
\normalsize
The overall optimization problem is thus formulated as:
\begin{align}
\small
    &\hspace{-0.1in} \max_{\mathbf{a}_w, r_m(t), \mathbf{Z}_w, R_s^w} \hspace{-0.2in}F_1\left(\mathbf{a}_w, r_m(t), \mathbf{Z}_w, R_s^w\right)-\nu F_2\left(r_m(t)\right), \label{obj} \\ 
    &\hspace{-0.05in} \text{s.t.} \hspace{0.18in} \sum_{s \in \mathcal{S}} R_{s}^{w} \leq \mathcal{R}, \quad \forall w \in \mathcal{W}, \label{c-1} \\
    & \hspace{0.22in} \sum_{m\in \mathcal{M}_s^{t}}r_m(t)\leq R_s^w, \quad \forall t\in \mathcal{T}_w, \label{c-2} \\
    & \hspace{0.3in} V_s\left(\mathbf{a}_w, r_m(t), \mathbf{Z}_w, R_s^w\right) \geq \epsilon^s, \quad \forall s \in \mathcal{S}, \label{c-3}\\
    &\hspace{0.3in}  \sum_{s\in \mathcal{S}}z_{m,s}^w=1, \quad \forall m \in \mathcal{M}_b^{t}, \quad\forall t\in \mathcal{T}_w, \label{c-4}
\end{align}
\normalsize
where $\nu$ is a scalar balancing the trade-off between performance and efficiency, and $\epsilon^s \in (0,1)$ is the minimum reliability threshold for slice $s$. 
The constraint in~\eqref{c-3} enforces the expected long-term reliability of each slice, defined as $V_s\left(\mathbf{a}_w, r_m(t), \mathbf{Z}_w, R_s^w\right)=\mathbb{E}\left[\Omega_s\left(.\right) + \sum_{t'=1}^{\infty} \zeta^{t'} \Omega_s\left(.\right)\right]$, with arguments of $\Omega_s(.)$ implicitly shifting over future windows as in~\eqref{obj-2}. 

\section{Methodology \& Approach} \label{section3}

To coordinate admission, association, and orchestration under dynamic conditions, we model the system as an MMDP and develop \textsc{SliceWise}, a DRL framework combining LSTM encoders with a D3QN for temporally aware control.


\subsection{MMDP Framework and Agent Design}

The proposed system consists of a set of agents $\mathcal{N}=\{\operatorname{ac},\operatorname{sa},\operatorname{re}_1,\operatorname{re}_2,\dots, \operatorname{re}_S\}$, where ``$\operatorname{ac}"$ denotes the admission control agent, ``$\operatorname{sa}"$ the slice association agent, and 
``$\operatorname{re}_s"$ the resource orchestrator for slice $s \in \mathcal{S}$. The MMDP is defined by the tuple $\left(\mathcal{N}, \mathcal{S_{\operatorname{env}}}, \{\mathcal{O}_i\}_{i\in\mathcal{N}}, \{\mathcal{A}_i\}_{i\in\mathcal{N}}, \mathbf{P}_{\operatorname{env}}, \{R_i\}_{i\in\mathcal{N}}, \gamma\right),$ where $\mathcal{S_{\operatorname{env}}}$ represents the global environment state, $\left(\mathcal{O}_i\right)$ and $\left(\mathcal{A}_i\right)$ are the observation and action spaces of agent $i \in \mathcal{N}$, $\mathbf{P}_{\operatorname{env}}$ is the transition model, $R_i$ is the reward function, and $\gamma \in (0,1)$ is the discount factor. Each agent follows a policy $\pi_i : \mathcal{O}_i \rightarrow \mathcal{A}_i$, selecting actions $\mathcal{A}_i \sim \pi_i(A_i | o_i)$ based on local observations $o_i$. Agents receive feedback from the environment through rewards $R_i(o_i,A_i)$, guiding policy updates over time.

\subsection{Observations and Action Spaces}

\begin{figure*}[t!]
    \centering
    \begin{subfigure}[b]{0.245\textwidth}\includegraphics[width=\textwidth]{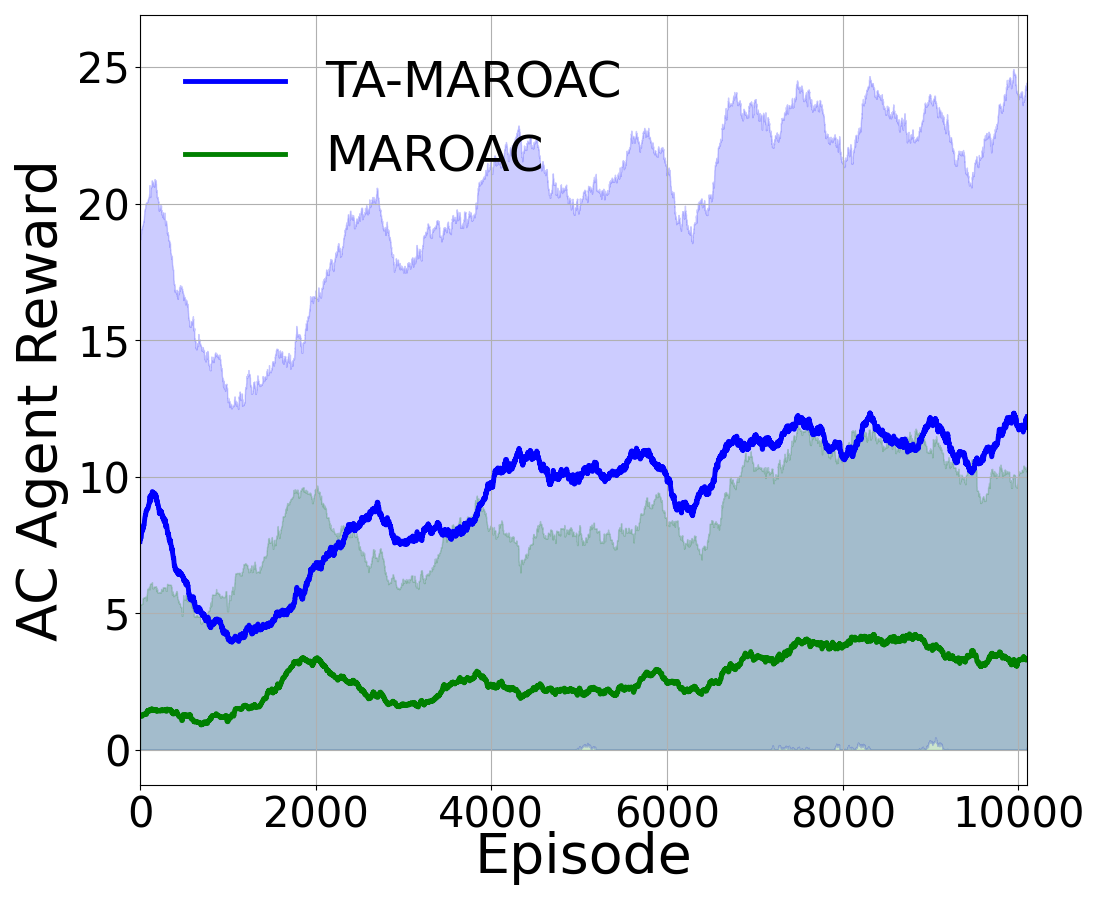}
        \caption{\vspace{-0.1in}}
    \end{subfigure}
    \hfill
    \begin{subfigure}[b]{0.245\textwidth}
        \includegraphics[width=\textwidth]{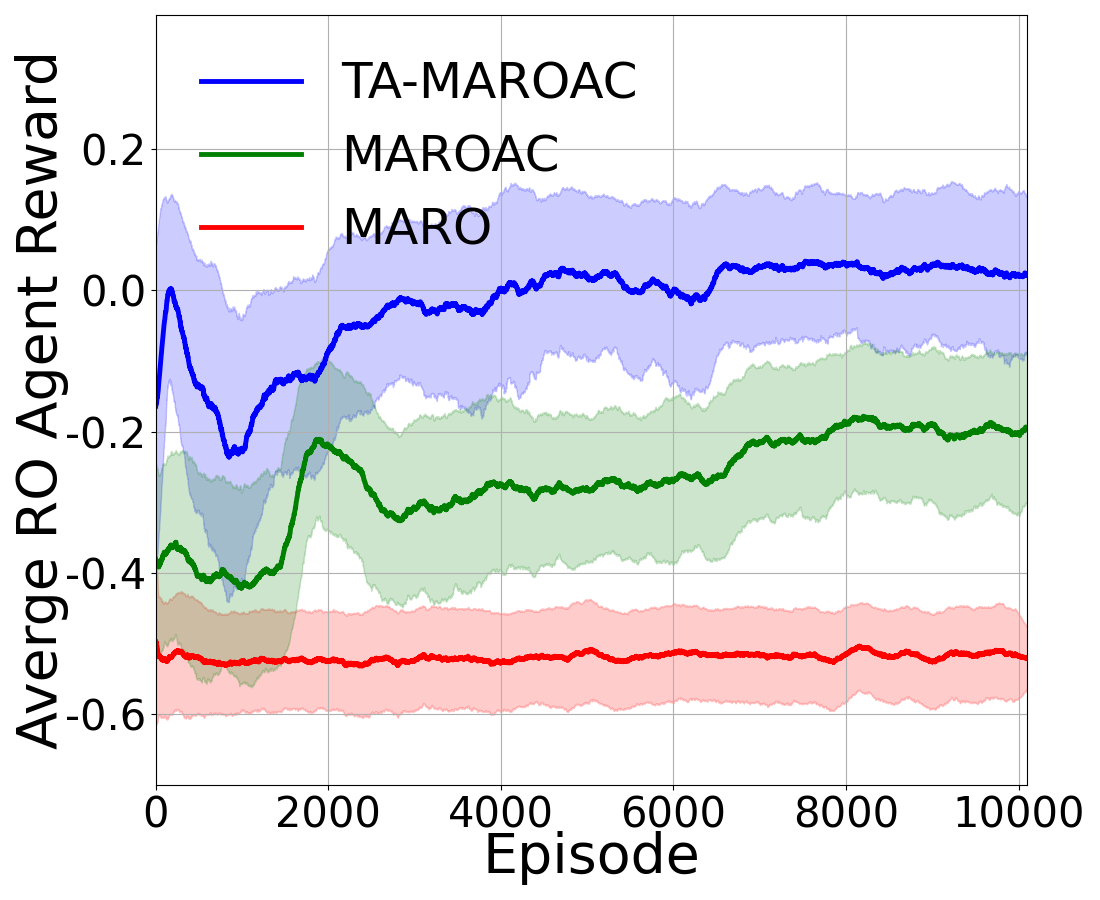}
        \caption{\vspace{-0.1in}}
    \end{subfigure}
    \hfill
    \begin{subfigure}[b]{0.245\textwidth}
        \includegraphics[width=\textwidth]{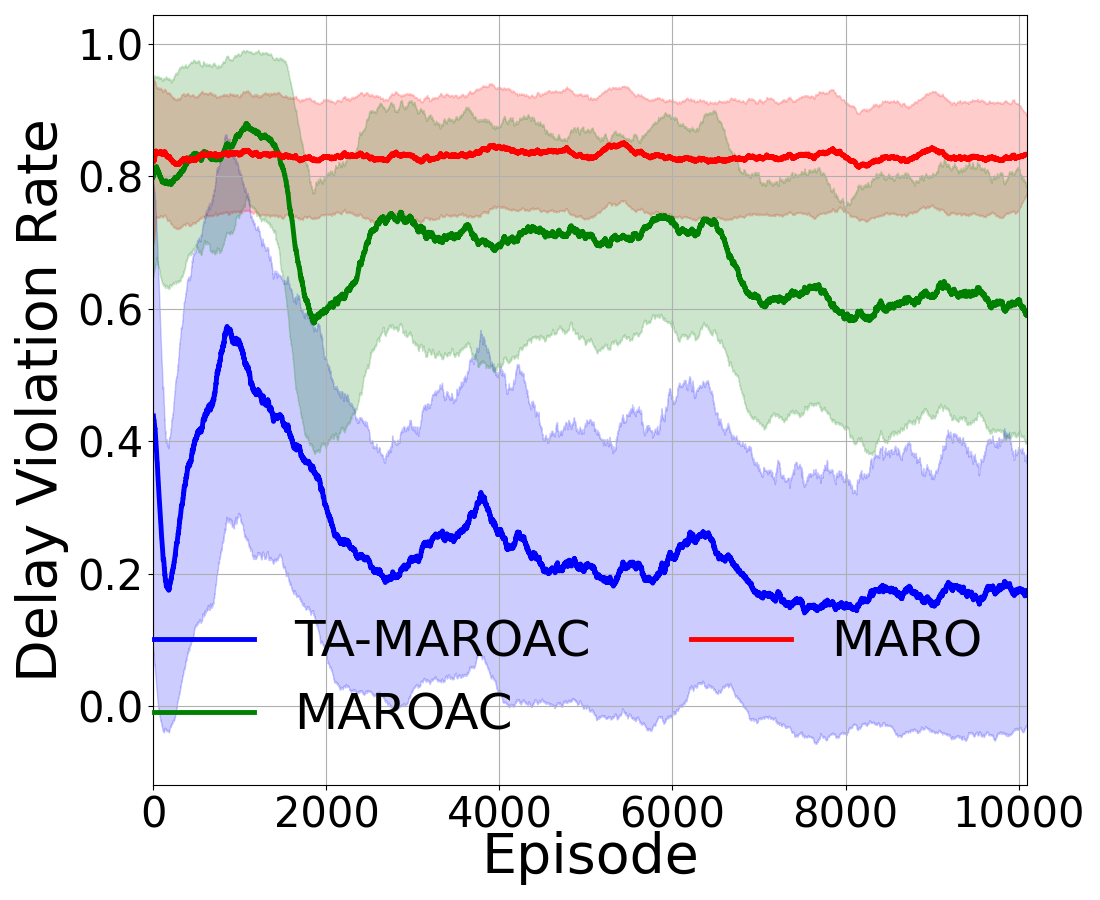}
        \caption{\vspace{-0.1in}}
    \end{subfigure}
    \begin{subfigure}[b]{0.245\textwidth}
        \includegraphics[width=\textwidth]{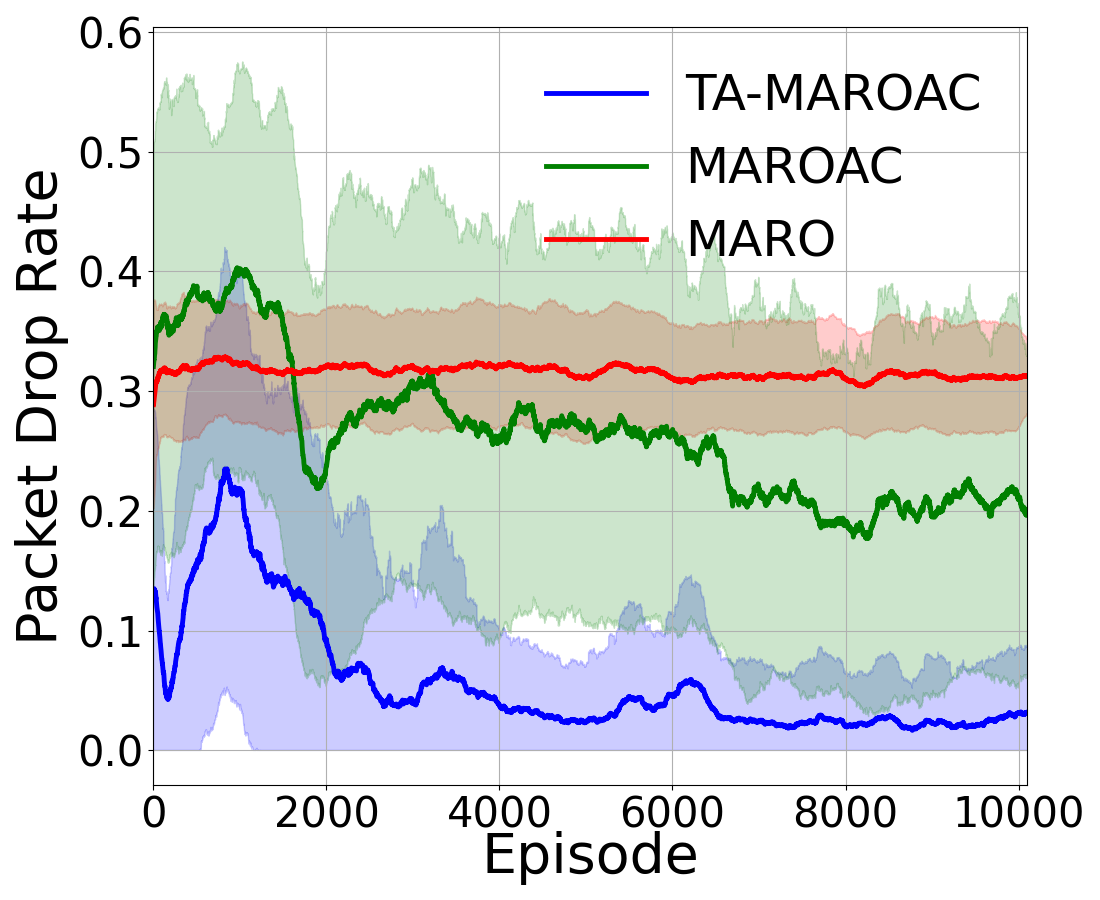}
        \caption{\vspace{-0.1in}}
    \end{subfigure}
    \caption{\small Evolution of training metrics – (a) the reward of admission control agent, (b) the reward of resource allocation agent, (c) the delay violation rate, and (d) packet drop rate. \vspace{-0.2in}}
    \label{fig:training_KPIs}
\end{figure*}

\noindent\textit{1) Admission Control Agent:} At the beginning of each time window $t_w\in \mathcal{T}_w$, the admission control agent observes a compact summary of the network's current load and quality conditions. This includes available bandwidth, recent delay-violation trends, aggregate queue statistics, and average channel quality across both admitted and requesting UEs. Formally, its observation vector is given by:

$
o_{\operatorname{ac}}^{t_w}=\bigg[\mathcal{R}\cdot B, \vartheta_{w-1}, \varrho_{w-1}, \varsigma_{w-1}, \frac{1}{M_a^{t_w-1}}\hspace{-0.15in}\underset{m \in \mathcal{M}_a^{t_w-1}}{\sum}\hspace{-0.15in}\tau_m^{\max}, M_a^{t_w-1}, \\ M_{b}^{t_w-1},
\underset{m \in \mathcal{M}_b^{t_w-1}}{\sum}\hspace{-0.15in}{r}_{m}(t_w-1)B, \frac{1}{M_a^{t_w-1}}\hspace{-0.15in}\underset{m \in \mathcal{M}_a^{t_w-1}}{\sum}\hspace{-0.2in}\operatorname{SINR}^{t_w-1}_m, \\ \frac{1}{M_b^{t_w-1}}\hspace{-0.15in}\underset{m \in \mathcal{M}_b^{t_w-1}}{\sum}\hspace{-0.15in}\hspace{-0.05in}\operatorname{SINR}^{t_w-1}_m\hspace{-0.05in},
 \frac{1}{M_b^{t_w-1}}\hspace{-0.15in} \underset{m \in \mathcal{M}_b^{t_w-1}}{\sum}\hspace{-0.15in}\tau_m^{\max}\bigg]\hspace{-0.01in},
$
where averaged system delay violation rate within a time window $\vartheta_{w}= \frac{1}{T_w} \underset{{t\in \mathcal{T}_{w}}}{\sum} \frac{|\{m \in \mathcal{M}_b^t \mid \mathbbm{1}_{\tau_m(t) \geq \tau_{m}^{\max}}\}|}{M_b^t}$. 
Similarly, the average data processing requests from admitted users and new applicants are respectively
$\varrho_{w} = \frac{1}{T_w}\hspace{-0.05in} \underset{{t\in \mathcal{T}_{w}}}{\sum} \hspace{-0.05in}\frac{1}{M_a^{t}}\hspace{-0.1in}\underset{m \in \mathcal{M}_a^{t}}{\sum} \hspace{-0.1in} N_m(t)\delta_m(t)$ and $ \varsigma_{w} = \frac{1}{T_w} \hspace{-0.05in}\underset{{t\in \mathcal{T}_{w}}}{\sum}\hspace{-0.05in}\frac{1}{M_b^{t}}\hspace{-0.1in}\underset{m \in \mathcal{M}_b^{t}}{\sum}\hspace{-0.1in}N(t)_m\delta_m(t)$.
To maintain system feasibility, the maximum number of admissible UEs per time slot is constrained to \( M_{\text{max}} \). Based on this observation, the agent selects an action \(A_{\operatorname{ac}} \in \mathcal{A}_{\operatorname{ac}} = \{1, 2, \dots, M_{\text{max}}\} \), where each action represents admitting a subset of candidate UEs according to the learned policy that balances SINR conditions, queue occupancy, and latency satisfaction.

\noindent\textit{2) Resource Orchestration Agent:} The orchestration agent operates at a finer timescale to adjust PRB allocations for each slice $s$ during every slot $t \in \mathcal{T}_w$. Its observation combines delay bounds, average SINR, traffic load, and current slice allocation: 
$
o_{\operatorname{re}_s}^t \hspace{-0.05in}=\hspace{-0.05in}\left[\frac{1}{M_s^{t-1}}\underset{m \in \mathcal{M}_s^{t-1}}{\sum}\tau_m^{\max}, \frac{1}{M_s^t}\underset{m \in \mathcal{M}_s^{t-1}}{\sum}\operatorname{SINR}^{t-1}_m,\right.\\
\hspace{0.25in}\left. \underset{m \in \mathcal{M}_s^{t-1}}{\sum} \hspace{-0.1in} r_{m}(t-1)B, \frac{1}{M_s^{t-1}} \hspace{-0.1in}\underset{m \in \mathcal{M}_s^{t-1}}{\sum}N_m(t-1)\delta_m(t-1), R_s^w B \right].
$ 
From this observation, the agent selects \( A_{\operatorname{re}_s}^t \in \mathcal{A}_{\operatorname{re}_s} = \{1, 2, \dots, N_{\operatorname{re}_s}\} \), corresponding to PRB adjustment factors $\rho_s^t$ that scale the slice-level allocation $R_s^{w}$. The resulting per-UE bandwidth update is:
\begin{align}
\small
    &r_{m}(t) = r_m(t-1) + \Big\lfloor\rho_s^t R_s^{w} \frac{\hat{B}_m(t)}{\sum_{m \in \mathcal{M}_s^t} \hat{B}_m(t)}\Big\rfloor, \ m \in \mathcal{M}_s^{t},
\end{align}
\normalsize
where $\hat{B}_m(t)=\frac{\delta_m(t) N_m(t)+\left(\tau_m^{\text{t}}(t-1)-T_s\right)^{+} b_m(t-1)}{\tau_m^{\textrm{max} }}$ estimates per-user bandwidth demand.
To ensure resource allocation remains within capacity limits, we employ the minimum adjustment method.

\noindent\textit{3) Slice Association Agent:} The association agent observes recent traffic sequences of all active users to infer their volatility patterns:
$ o_{\operatorname{sa}}^{t_w}=\big[D_1^{w}, D_2^{w},\dots, D_{M_b^{t_w}}^{w}\big],
$
where, $D_m^{w} = \big[N(t_w-1)_m\delta_m(t_w-1), \dots, N^{(t_w-\theta T_w)}_m\delta_m^{(t_w-\theta T_w)}\big]^{\text{T}}$.
The parameter $\theta$ defines the number of past time windows considered, thereby controlling the temporal horizon over which traffic variability is captured. The resulting association matrix 
$\mathbf{Z}_w$ is defined as in Section~\ref{section2}.

\subsection{Reward Functions}
The admission control agent's reward function is formulated to optimize user admissions while ensuring stringent reliability constraints, which mathematically is defined as follows:
\begin{align}
\small
\hspace{-0.12in}R_{\operatorname{ac}}^{t_w}= \hspace{-0.03in}u \frac{M_k^{{t_w}}}{M_a^{t_w-1}} \hspace{-0.05in}\prod_{m \in \mathcal{M}_b^{t}}\hspace{-0.05in} G_{c}\left( \tau_m^{\max} - \bar{\tau}_m^{\text{t}}(t_w)\right),
\end{align}
\normalsize
where function $G_{c}(x)\hspace{-0.03in}= \hspace{-0.03in}\frac{1}{1 + \exp\left(-(\alpha_c(x +\varphi_c)\right)}$. Average transmission delay at time slot $t_w$ is calculated as $\bar{\tau}_m^{\text{t}}(t_w)=\frac{1}{T_w}\underset{t\in\mathcal{T}_w}{\sum}\tau_m^{\text{t}}(t)$ and $u$ is a scaling factor. For the resource agent in charge of the orchestration of slice $s$, the reward at time slot $t\in \mathcal{T}_w$ reflects system performance in terms of delay violation and resource efficiency defined as:
\begin{align}
\small
R_{\operatorname{re}_s}^t \hspace{-0.05in}=\hspace{-0.03in} \frac{1}{M_s^t} \hspace{-0.03in} \sum_{m \in \mathcal{M}_s^{t}} \hspace{-0.05in} G_s \hspace{-0.03in}\left( \tau_{m}^{\max} \hspace{-0.03in} -\tau_m^{\text{t}}(t)\right) \hspace{-0.03in} -\nu G_u\Big(\frac{1}{ R_s^w} \hspace{-0.05in} \sum_{m \in  \mathcal{M}_{s}^t} \hspace{-0.06in} r_m(t) \Big),
\end{align}
\normalsize
where $G_s(x)=2g_s^r\hspace{-0.04in}\left( x\right)-1$ and $G_{u}(x)\hspace{-0.03in}= \hspace{-0.03in}\frac{1}{1 + \exp\left(-\alpha_u (x +\varphi_u)\right)}$.

\subsection{Learning and Training Strategy}

Training proceeds over multiple episodes of $T$ time slots. At episode start, UEs are randomly assigned eMBB or URLLC service types with corresponding delay constraints. Traffic dynamics, including packet count $N_m(t)$, packet size $\delta_m(t)$, and distance $d_m(t)$, evolve via Markov chains with transition matrices $\mathbf{P}^q_{\sigma}$, where $q\in\{1,2,3\}$ indexes variable type and $\sigma$ the traffic profile. Transitions are restricted to adjacent states to ensure smooth variability.

Each D3QN-based agent maintains separate target and online networks, with experience replay buffers of size $S^{\text{buffer}}_{\text{d3qn}}$ storing tuples $(o_i,A_i,R_i,o_i')$. Parameters include learning rate $\mu_{\text{d3qn}}$, discount factor $\gamma$, soft update factor $\lambda_{\text{d3qn}}^{\text{soft}}$, and batch size $S_{\text{d3qn}}^{\text{Batch}}$. Admission and association agents update at each reconfiguration window, while resource agents update every slot. An $\epsilon$-greedy policy with decaying $\epsilon$ guides exploration.

User–slice association is implemented as an LSTM classifier trained offline on traffic sequences $N_m(t)\delta_m(t)$ with volatility\footnote{In this work, \emph{volatility} denotes the degree of temporal variation in UE’s traffic, characterized by fluctuations in packet arrival count and packet size.} labels and kept fixed during RL. This preserves a stable mapping from temporal traffic patterns to slice labels, reduces non-stationarity in the control loop, and meets near-RT inference constraints.
At runtime, the LSTM outputs slice mappings, followed by inter-slice resource allocation, i.e., 
$
R_s^w = \left\lfloor \frac{\mathcal{R}}{\sum_{s\in\mathcal{S}} M_s^{t_w}} \right\rfloor.
$
A warm-up period of $\theta T_w$ slots precedes policy execution.

\begin{table}[t]
\centering
\caption{\label{tab:SOTA} Simulation Parameters. \vspace{-0.0in}}
\footnotesize
\renewcommand{\arraystretch}{1.1} 
\begin{tabularx}{0.48\textwidth}
{
| >{\centering\arraybackslash}X
| >{\centering\arraybackslash}X
| >{\centering\arraybackslash}X
| >{\centering\arraybackslash}X |
}
\hline \textbf{Parameters} & \textbf{Value} & \textbf{Parameters} & \textbf{Value} \\
\hline $d_m(t)$ (m) & [400, 500] & $M$, $S$ & 15, 4 \\
\hline $N_m(t)$ & [50, 75] & $\delta_m(t)$ (Mb) & [0.003, 0.6] \\
\hline $N_0$ (dBm/Hz) & -163 & $\eta$, $\psi_{\xi}$ (dB) & 1.01, 2.23 \\
\hline $\tau_{\text{u}}$ (s) & 0.001 & $\tau_{\text{e}}$ (s) & 0.1 \\
\hline
$\gamma$ & 0.99 & $\mu_{\operatorname{d3qn}}$ & $5\times10^{-5}$ \\
\hline $u$ & 1000 & $N_{\text{train}}$ & 10000 \\
\hline $H^{\text{net}}_{\operatorname{d3qn}}$ & $\{25, 25\}$ & $N_{\text{test}}$ & 150 \\
\hline
$S^{\text{buffer}}_{\operatorname{d3qn}}$ & $5 \times 10^{5}$ & $\lambda^{\text{soft}}_{\operatorname{d3qn}}$ & $5 \times 10^{-3}$ \\
\hline $N_s^{1}, N_s^{2}, N_s^{3}$ & 10, 10, 10 & $\theta$, $\nu$& 5, 0.2 \\
\hline$\alpha_1^r, \alpha_2^r, \alpha_3^r,\alpha_4^r$& \mbox{33, 25, 20, 16} & $\varphi$& $-0.0005$\\
\hline$\alpha_c$, $\alpha_u$& 20, 10&$\varphi_c$, $\varphi_u$ &$-0.0005, -0.6$\\
\hline
 $H^{\text{net}}_{\operatorname{lstm}}$ &\{64, 64, 64\} & $\mu_{\operatorname{lstm}}$ & 0.0005 \\
\hline
$S^{\operatorname{Batch}}_{\operatorname{d3qn}}$ & 256 & $S^{\operatorname{Batch}}_{\operatorname{lstm}}$ & 256 \\
\hline
\end{tabularx}
\end{table}

\section{Simulation Results and Analysis} \label{section4}

The proposed \textsc{SliceWise} framework is implemented and evaluated on a customized Python-based version of the open-source \texttt{AI-RAN Simulator}~\cite{ai-ransim}. 
The setup emulates a multi-cell O-RAN with a serving O-RU and two interfering O-RUs positioned 2.5 and 3~km apart, all operating on the same 2.5~GHz carrier. Inter-cell interference is incorporated in each UE’s SINR.
The total bandwidth under carrier aggregation $B_{\text{total}} \in \{350, \dots, 1500\}$~MHz is discretized into $\mathcal{R} \approx \lfloor B_{\text{total}} / (B_{\text{scs}} \times 12) \rfloor$ PRBs with subcarrier spacing $B_{\text{scs}} = 15$~kHz. Key parameters and learning hyperparameters appear in Table~\ref{tab:SOTA}.
Each simulation episode consists of $T=100$ time slots of duration $T_s = 0.2$~s. Training is conducted for $N_{\text{train}} = 10{,}000$ episodes, where the orchestration agent acts at every slot while admission and slice-association agents update once every $T_w = 20$ slots. Results are averaged over $N_{\text{test}} = 150$ randomized test episodes.

The network accommodates $S = 4$ active slices and $M = 15$ UEs distributed among them. Each UE’s traffic process is governed by a first-order Markov chain, capturing temporal correlation in both packet arrival count $N_m(t)$ and packet size $\delta_m(t)$. For each slice, the instantaneous values of these variables are drawn uniformly from pre-defined intervals, with transitions constrained by a probability of~0.5, allowing either self-transition or progression to the next volatility state. To systematically examine framework adaptability, slices are configured with escalating volatility levels.

Slice~1 (benign) operates with narrow, stable ranges: for eMBB, $\delta_m(t)\!\in\![0.01, 0.02]$~MB and $N_m(t)\!\in\![80, 240]$; for URLLC, $\delta_m(t)\!\in\![0.256, 0.32]$~Kb and $N_m(t)\!\in\![200, 400]$. This variability is incrementally expanded across slices. For example, Slice~4 (highly volatile) uses $\delta_m(t)\!\in\![0.01, 0.06]$~MB and $N_m(t)\!\in\![80, 720]$ for eMBB, and $\delta_m(t)\!\in\![0.256, 0.64]$~Kb with $N_m(t)\!\in\![200, 1600]$ for URLLC. While traditional slice assignment relies on predefined QoS classes, we infer association dynamically from observed traffic volatility, enabling adaptive alignment between user behavior and slice characteristics. 

\begin{figure}[t!]
    \centering
    \begin{subfigure}[b]{0.235\textwidth}\includegraphics[width=\textwidth]{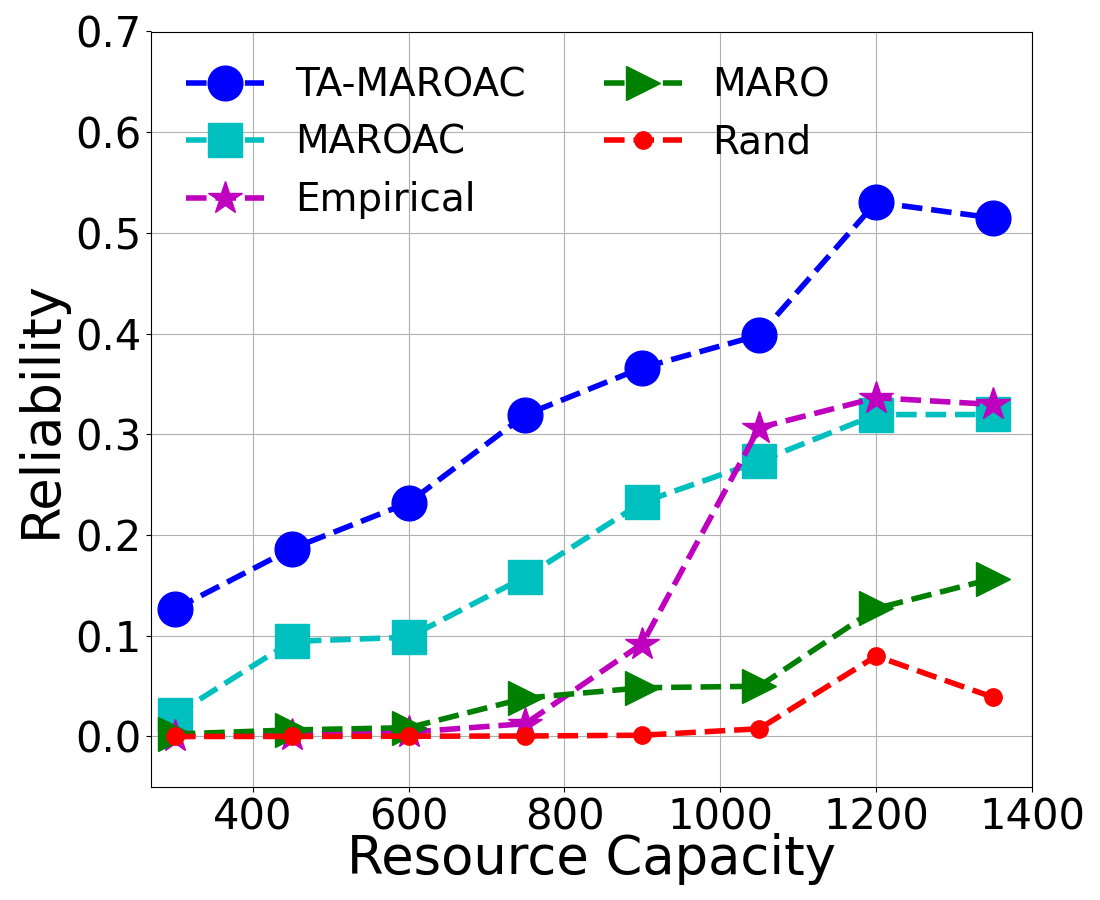}
        \caption{\vspace{-0.0in}}
    \end{subfigure}
    \hfill
    \begin{subfigure}[b]{0.235\textwidth}
        \includegraphics[width=\textwidth]{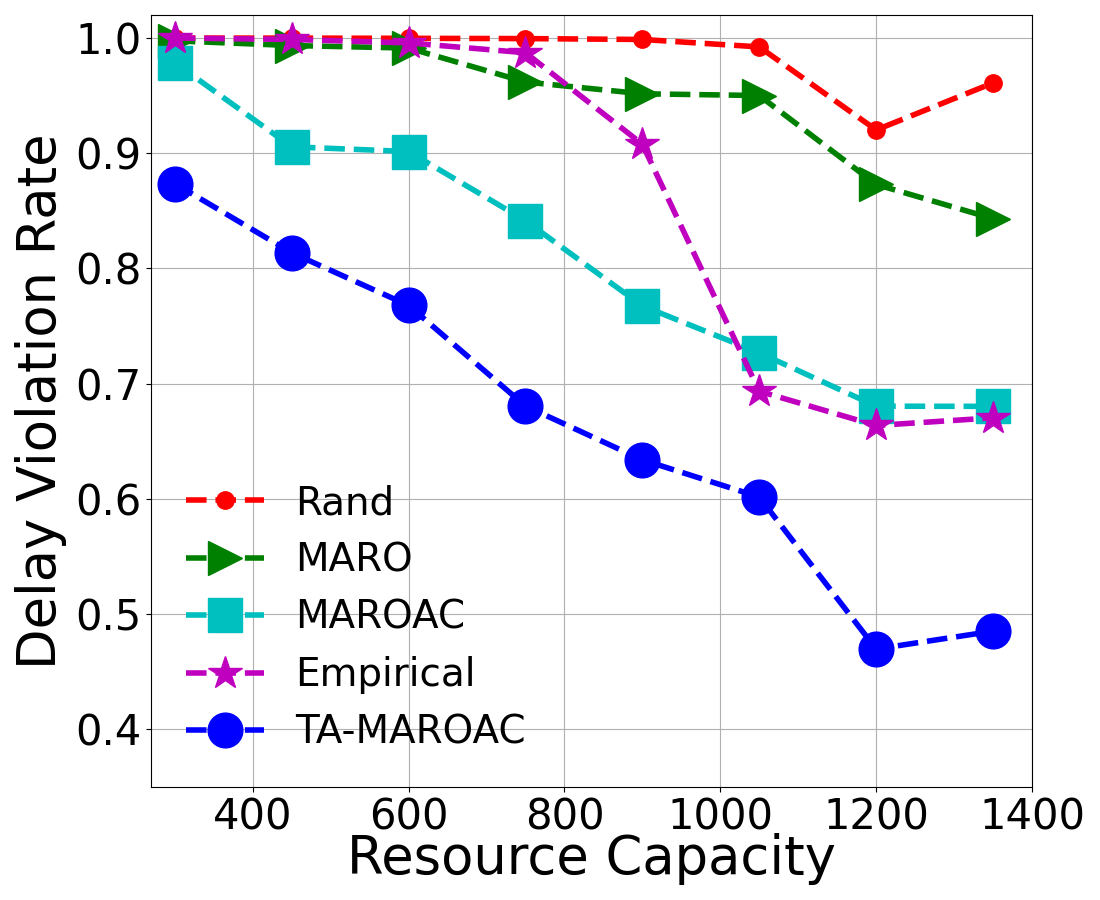}
        \caption{\vspace{-0.0in}}
    \end{subfigure}
    \hfill
    \begin{subfigure}[b]{0.235\textwidth}
        \includegraphics[width=\textwidth]{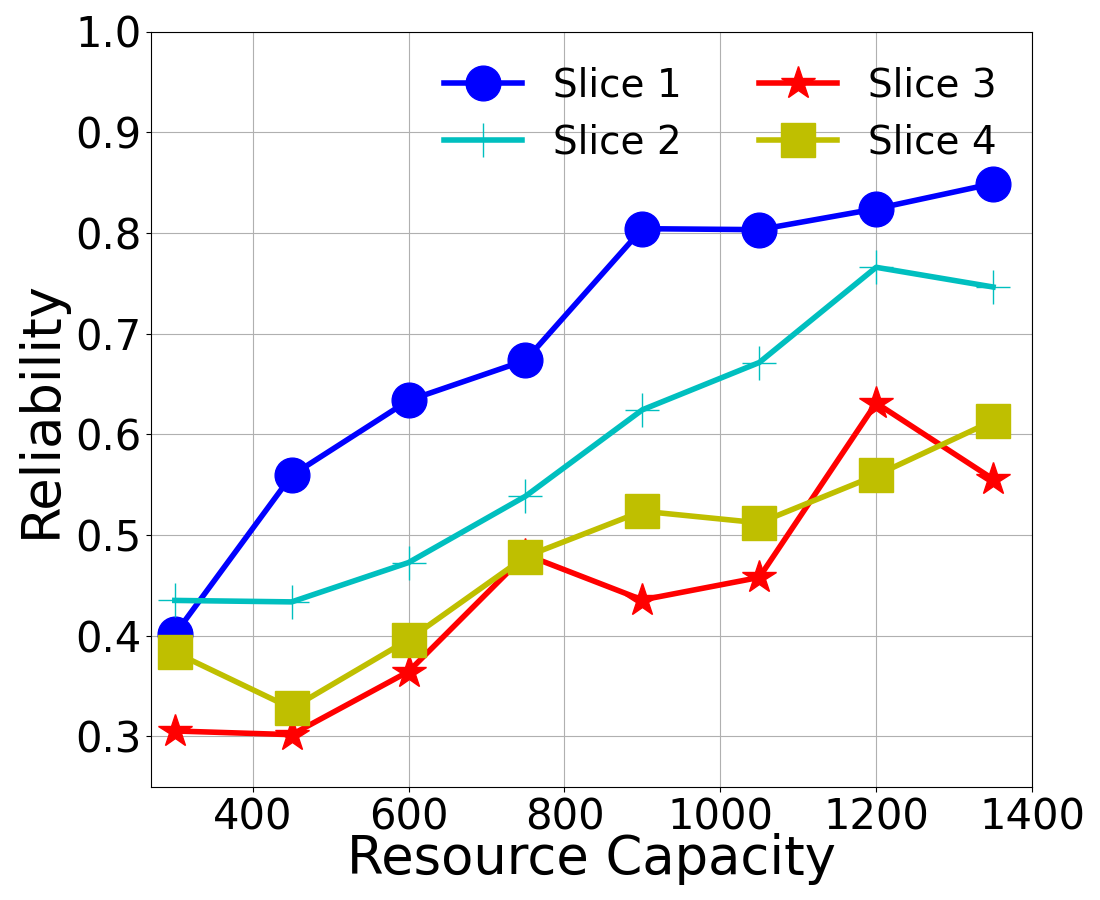}
        \caption{\vspace{-0.1in}}
    \end{subfigure}
    \begin{subfigure}[b]{0.235\textwidth}
        \includegraphics[width=\textwidth]{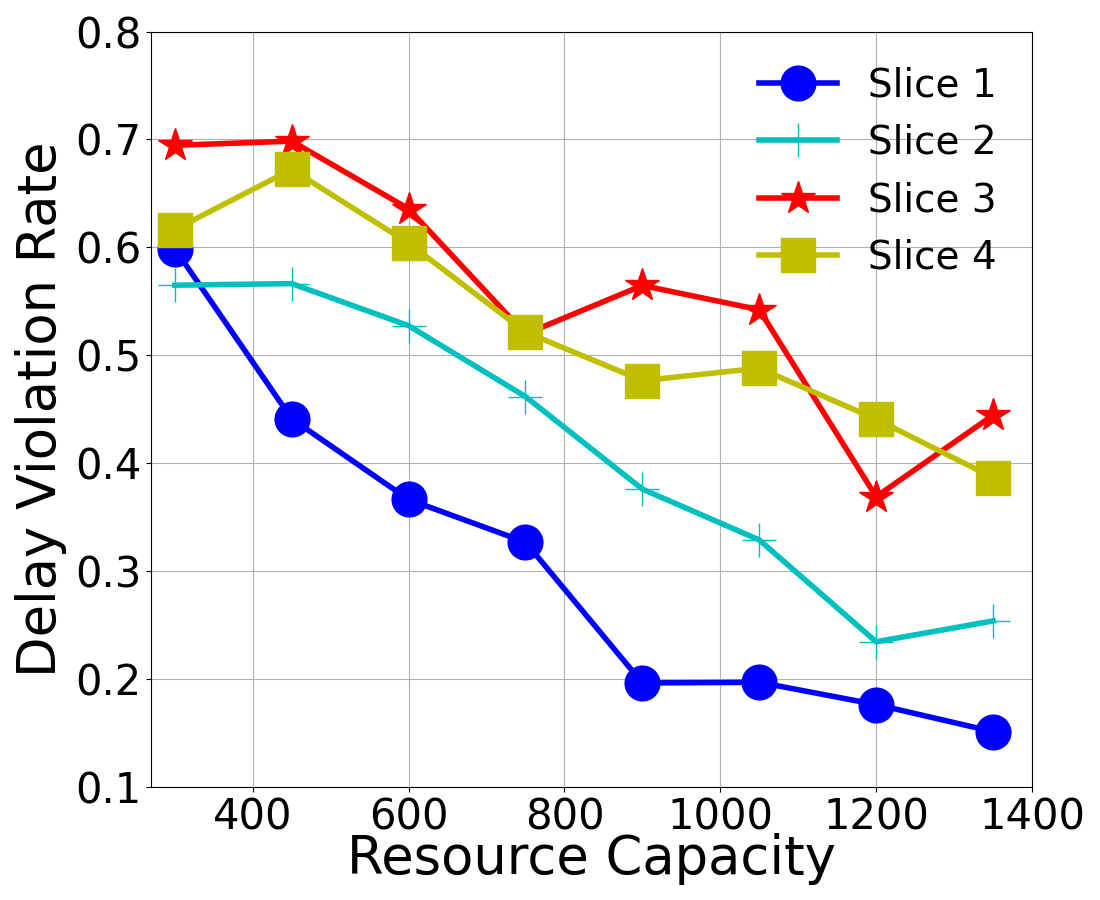}
        \caption{\vspace{-0.1in}}
    \end{subfigure}
    \caption{Comparative evaluation results varying $B_{\text{total}}$ in MHz – (a) system reliability, (b) system delay violation (c) silce-wise reliability, and (d) slice-wise delay violation.\vspace{-0.0in}}
    \label{fig:evaluations-1}
\end{figure}

As shown in Figs.~\ref{fig:training_KPIs}(a)–(b), cumulative rewards for admission and orchestration agents increase monotonically, indicating stable convergence. Delay-violation and packet-drop rates (Figs.~\ref{fig:training_KPIs}(c)–(d)) decline rapidly, with the traffic-aware configuration (\texttt{TA-MAROAC}) converging faster and outperforming \texttt{MARO} and \texttt{MAROAC}, confirming the advantage of temporal modeling and joint optimization.
The trained policy is tested on unseen multi-cell episodes with variable bandwidth and heterogeneous traffic. Compared with \texttt{Rand}, \texttt{Empirical}, \texttt{MARO}, and \texttt{MAROAC}, \textsc{SliceWise} achieves the highest reliability and lowest delay violations (Figs.~\ref{fig:evaluations-1}(a)–(b)). Slice-level results (Figs.~\ref{fig:evaluations-1}(c)–(d)) show sustained URLLC latency compliance even under interference.

\begin{figure}[h]
    \centering
    \begin{subfigure}[b]{0.237\textwidth}\includegraphics[width=\textwidth]{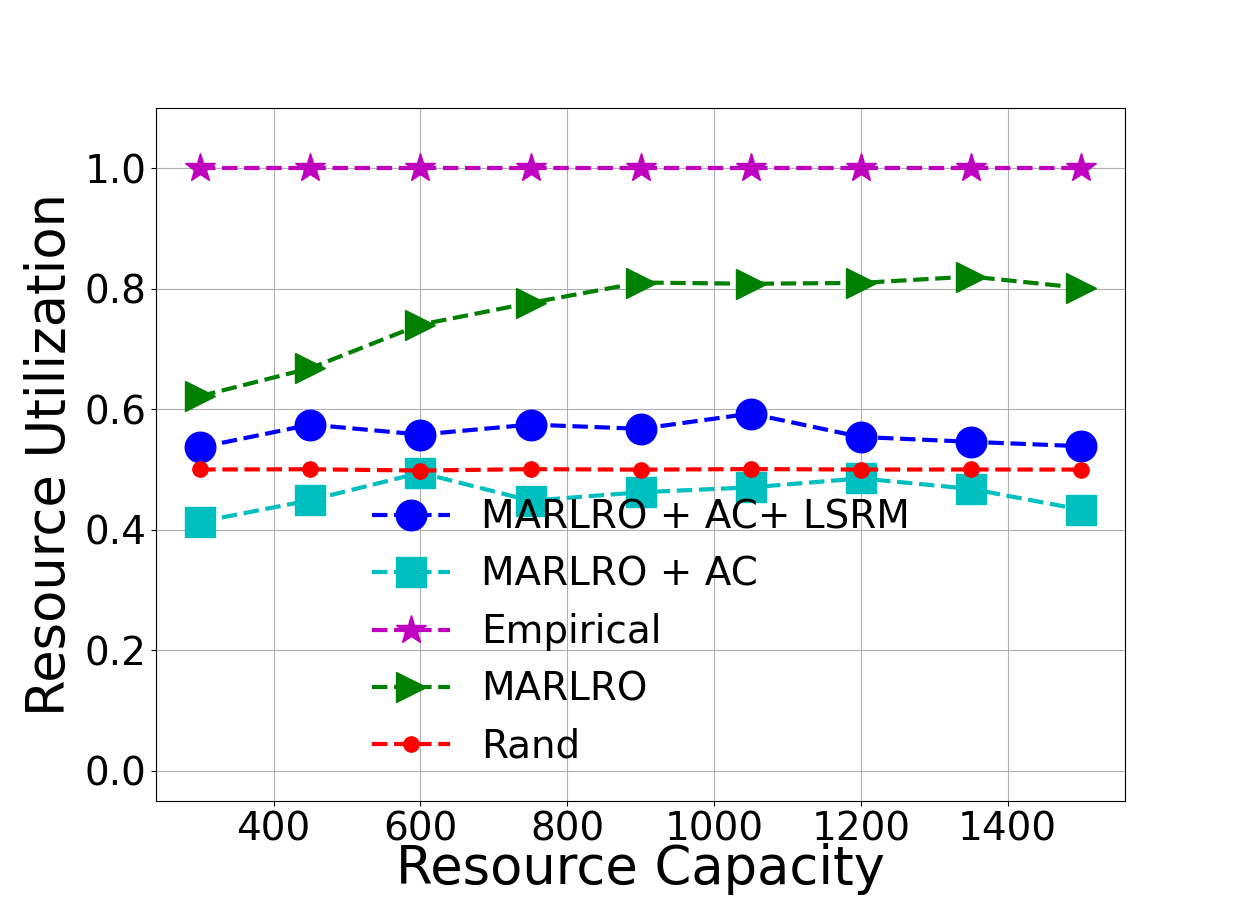}
        \caption{\vspace{-0.15in}}
    \end{subfigure}
    \hfill
    \begin{subfigure}[b]{0.237\textwidth}
        \includegraphics[width=\textwidth]{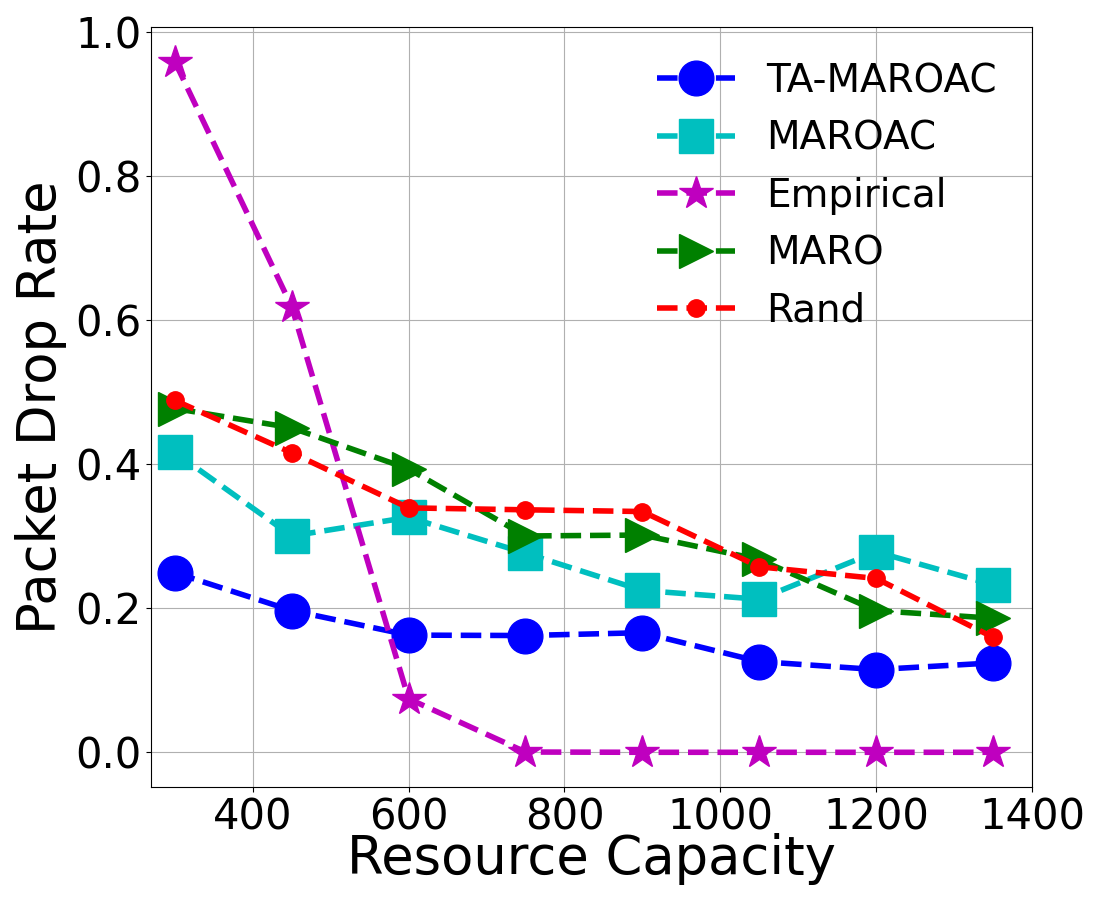}
        \caption{\vspace{-0.15in}}
    \end{subfigure} 
    \vskip\baselineskip
    \begin{subfigure}[b]{0.237\textwidth}
        \includegraphics[width=\textwidth]{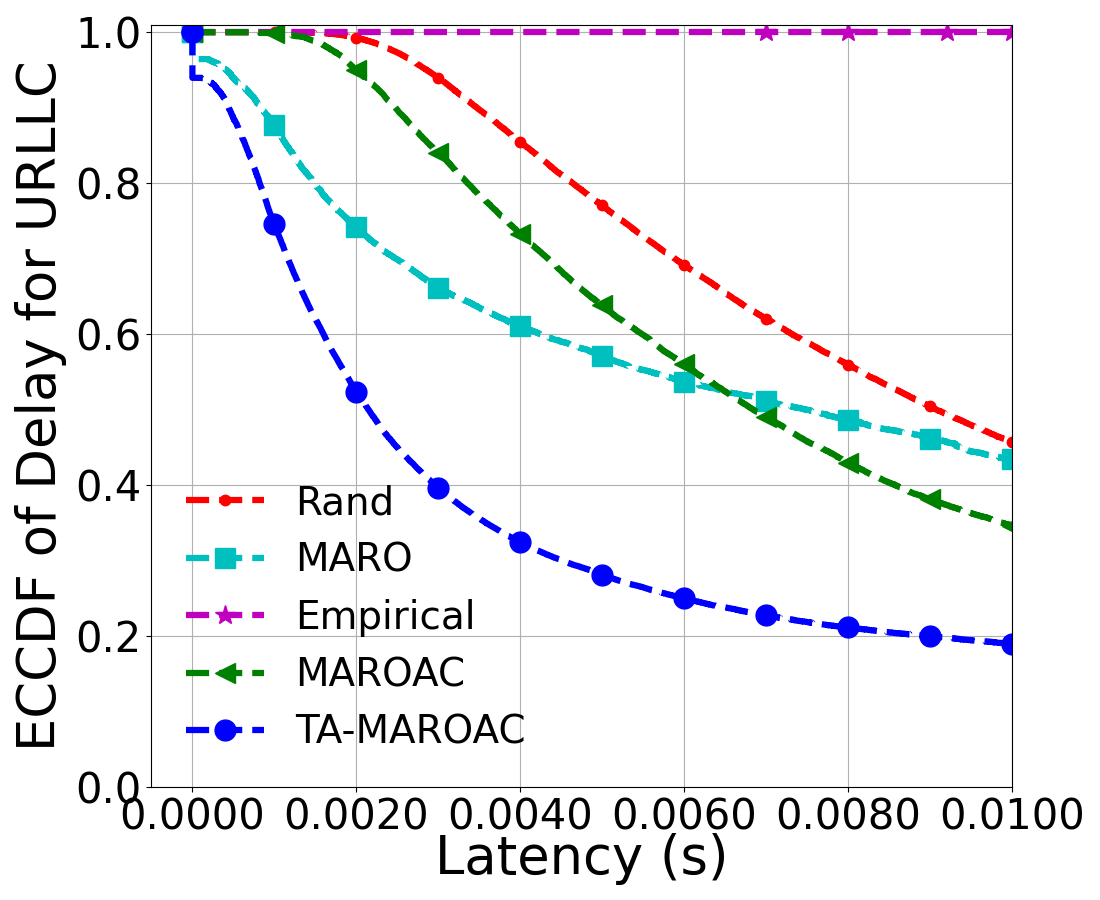}
        \caption{\vspace{-0.1in}}
    \end{subfigure}
    \begin{subfigure}[b]{0.237\textwidth}
        \includegraphics[width=\textwidth]{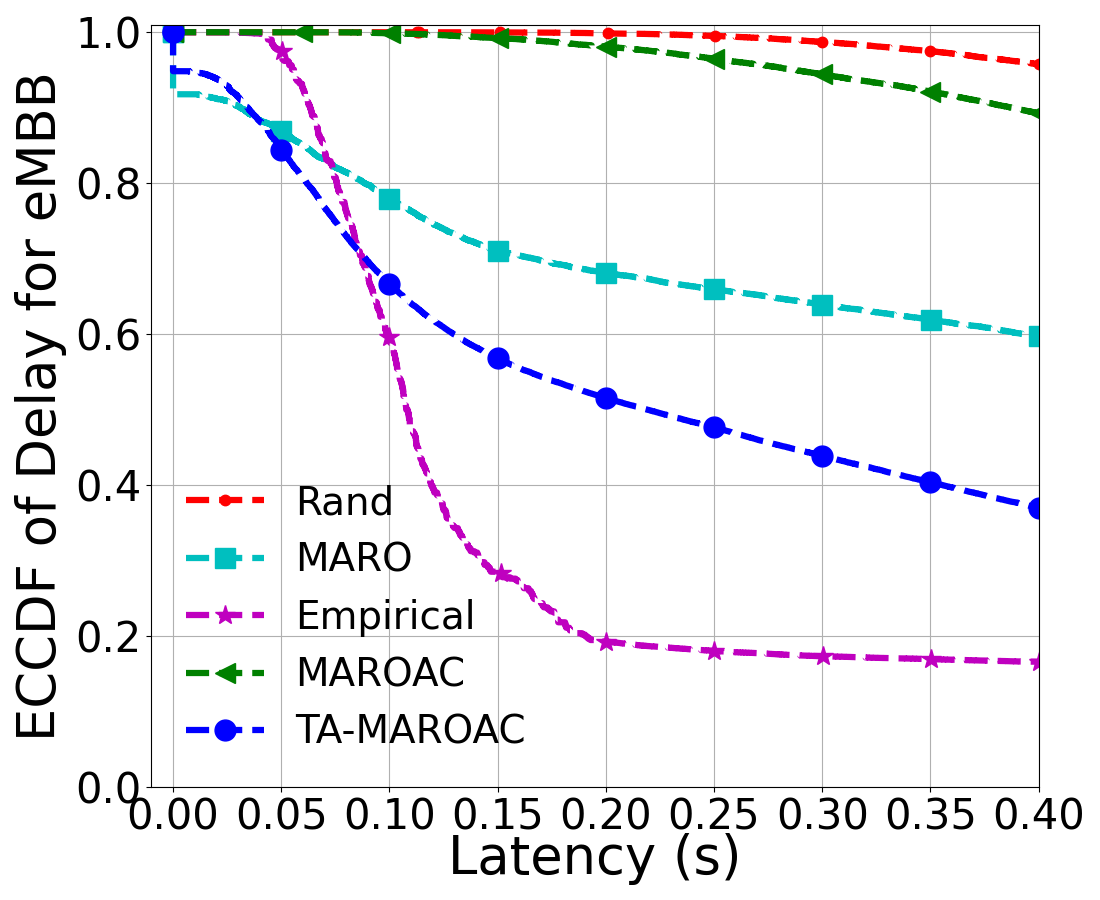}
        \caption{\vspace{-0.1in}}
    \end{subfigure}
    \caption{Comparative evaluation results varing $B_{\text{total}}$ in MHz – (a) system resource utilization, (b) system packet drop rate (c) ECCDF of delay of URLLC users, and (d) ECCDF of delay of eMBB users.\vspace{-0.0in}}
    \label{fig:evaluations-2}
\end{figure}

\begin{figure}[t!]
    \begin{subfigure}[b]{0.238\textwidth}
        \includegraphics[width=\textwidth]{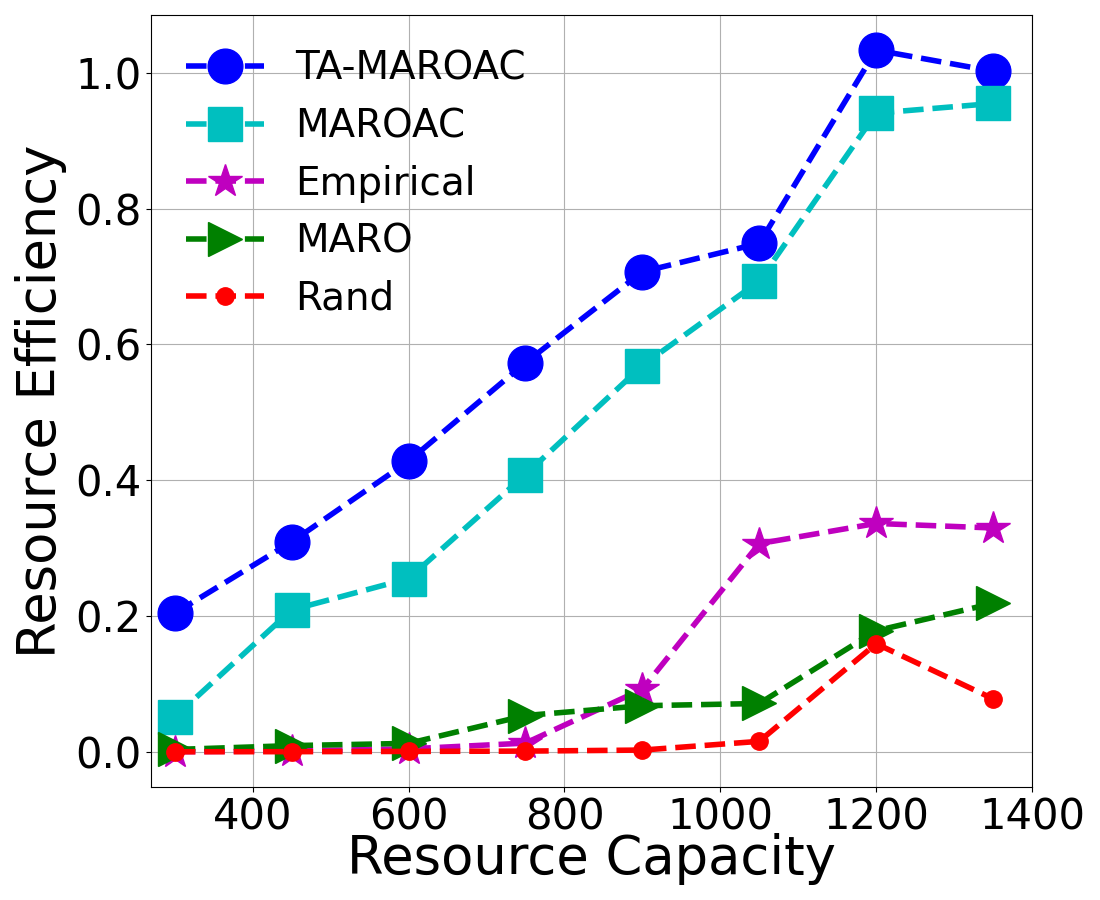}
        \caption{\vspace{-0.1in}}
    \end{subfigure}
    \begin{subfigure}[b]{0.238\textwidth}
        \includegraphics[width=\textwidth]{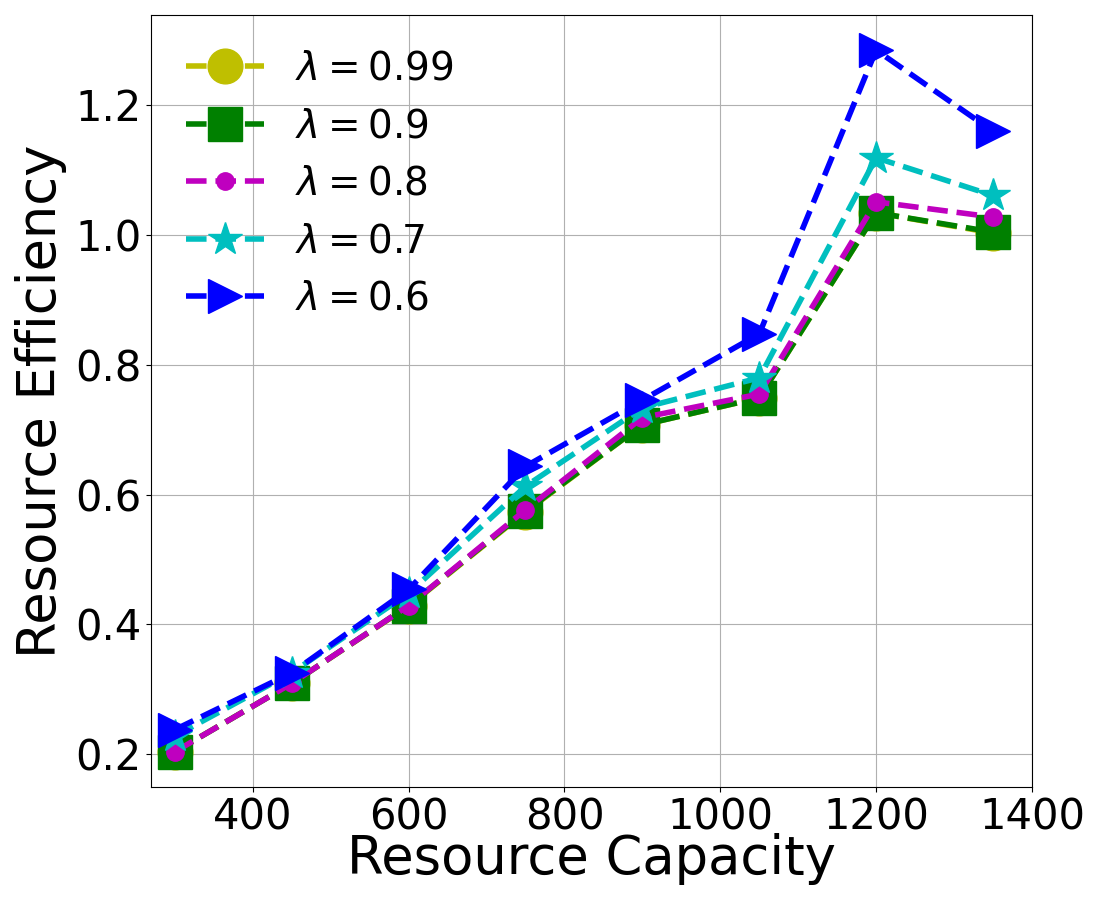}
        \caption{\vspace{-0.1in}}
    \end{subfigure}
    \caption{Resource efficiency evaluation varing $B_{\text{total}}$ in MHz-(a) resource efficiency under various orchestration schemes, (b) resource efficiency of the proposed scheme varing fulfillment threshold $\lambda$.\vspace{-0.05in}}
    \label{fig:resource-eff}
\end{figure}

Fig.~\ref{fig:evaluations-2}(a) demonstrates that \textsc{SliceWise} maintains efficient PRB utilization without over-provisioning. Although the \texttt{Empirical} scheme slightly reduces packet-drop rates (Fig.~\ref{fig:evaluations-2}(b)), its ECCDF delay profiles in Figs.~\ref{fig:evaluations-2}(c)–(d) exhibit long tails, indicating SLA violations. In contrast, \textsc{SliceWise} yields compact delay distributions for both eMBB and URLLC slices, reflecting stable queue dynamics and superior control over latency dispersion.
As shown in Fig.~\ref{fig:resource-eff}(a), \textsc{SliceWise} achieves the best joint optimization of user admission, SLA fulfillment, and resource utilization. The sensitivity analysis with respect to the latency fulfillment threshold $\lambda$ (Fig.~\ref{fig:resource-eff}(b)) shows that all methods degrade under stricter delay requirements, but \textsc{SliceWise} experiences slowest decline showing robustness and adaptability across varying conditions.

\vspace{-0.01in}
\section{Conclusion} \label{section5}
This paper introduced \textsc{SliceWise} a traffic-aware multi-agent reinforcement learning framework that seamlessly integrates resource orchestration and admission control within O-RAN environments. To address the strict slice-specific requirements of URLLC and eMBB traffic, our approach combines a multi-agent deep reinforcement learning model with an LSTM network to dynamically manage user admissions, slice associations, and resource allocations. This dual-model strategy enables real-time optimization of latency, reliability, and resource efficiency while effectively balancing multiple objectives amid evolving network conditions. Notably, the D3QN-based agents in our framework exhibit robust performance, achieving significant reductions in delay violation and packet drop rates compared to baseline methods.

\vspace{-0.00in}
\bibliographystyle{IEEEtran}
\bibliography{Bibliography}

\end{document}